# Visual Comfort Assessment for Stereoscopic Image Retargeting

Ya Zhou
University of Science and Technology of China
Hefei, China
zhouya@mail.ustc.edu.cn

Wei Zhou
University of Science and Technology of China
Hefei, China
weichou@mail.ustc.edu.cn

Ping An
Shanghai University
Shanghai, China
anping@shu.edu.cn

Zhibo Chen
University of Science and Technology of China
Hefei, China
chenzhibo@ustc.edu.cn

*Abstract*—In recent years, visual comfort assessment (VCA) for 3D/stereoscopic content has aroused extensive attention. However, much less work has been done on the perceptual evaluation of stereoscopic image retargeting. In this paper, we first build a Stereoscopic Image Retargeting Database (SIRD), which contains source images and retargeted images produced by four typical stereoscopic retargeting methods. Then, the subjective experiment is conducted to assess four aspects of visual distortion, i.e. visual comfort, image quality, depth quality and the overall quality. Furthermore, we propose a Visual Comfort Assessment metric for Stereoscopic Image Retargeting (VCA-SIR). Based on the characteristics of stereoscopic retargeted images, the proposed model introduces novel features like disparity range, boundary disparity as well as disparity intensity distribution into the assessment model. Experimental results demonstrate that VCA-SIR can achieve high consistency with subjective perception.

## I. INTRODUCTION

During the last few decades, 3D/stereoscopic technologies and applications have been widely used in various multimedia scenarios [1]. Compared with the quality assessment of traditional 2D image/video, 3D quality assessment is more sophisticated which involves multiple quality dimensions such as visual comfort, image quality, and depth quality [2]. Therefore, the visual comfort of stereoscopic images is one of the important aspects when viewing 3D images. Massive visual comfort assessment (VCA) models for stereoscopic images have been proposed. For example, in [3], relative disparity and object thickness features were developed to assess the visual comfort of 3D images. Also, [4] considered depth of focus, spatial frequency, and the zone of comfort related impacts. In [5] and [6], disparity magnitude, disparity contrast, disparity dispersion and disparity skewness features were introduced, which all belong to disparity statistics features. Besides, [6] also took the neural activities of the middle temporal area into account. However, the impacts of image distortion and binocular rivalry generally existing in stereoscopic retargeting scenarios have not been considered in above-mentioned studies.

At the same time, stereoscopic image retargeting technology has been studied to adapt stereoscopic images for 3D display devices with heterogeneous screen resolutions. Although many 2D image retargeting methods have been illustrated in [3][8]. However, there are fewer 3D image retargeting methods such as stereo cropping, stereo seam carving, stereo scaling and stereo multi-operator. Nevertheless, only image quality has been considered in existing quality assessment models for stereoscopic image retargeting to date [9][10]. In other words, these studies hardly consider the visual comfort from the angle of human perception, and lack of public 3D image retargeting quality assessment dataset is also one severe issue in this research area.

Therefore, in this paper, we first create a Stereoscopic Image Retargeting Database (SIRD) to study VCA for stereoscopic image retargeting. Specifically, this database consists of 100 source images and 400 retargeted images generated by stereo cropping, stereo seam carving, stereo scaling and stereo multi-operator four stereoscopic retargeting methods. Moreover, we propose an objective assessment algorithm named Visual Comfort Assessment metric for Stereoscopic Image Retargeting (VCA-SIR), which considers image distortion and disparity distribution. Finally, we tested VCA-SIR on our database and the results show that VCA-SIR has a wonderful visual comfort prediction performance, which can reach Pearson Linear Correlation Coefficient (PLCC) of 0.9024 in the case of full-reference and 0.8670 in the case of no-reference. Therefore, the proposed model can correlate well with human perception and should be applicable for stereoscopic image retargeting quality assessment under the scenarios of displaying 3D contents over heterogeneous devices.

For the other parts of this paper, we describe the details of establishing the stereoscopic image retargeting subjective database in Section II. In Section III, we introduce the novel image features designed in our proposed VCA-SIR. Then, the analysis of VCA-SIR's performance is described in Section IV. Finally, our conclusions are presented in Section V.

## II. STEREOSCOPIC IMAGE RETARGETING DATABASE

We build a subjective assessment database named SIRD for public research, using four various stereoscopic retargeting methods to evaluate the performance of VCA metrics.

### A. Selection of Stereoscopic Retargeted Images

One hundred source stereoscopic images with diversified properties are selected from IEEE-SA database [11] and IVY database [12]. Specifically, the property distribution of these images is shown in Table I. From Table I, we can see that the

This work was supported in part by the National Key Research and Development Program of China under Grant No. 2016YFC0801001, NSFC under Grant 61571413, 61632001, 61390514, and Intel ICRI MNC.

TABLE I. THE PROPERTY DISTRIBUTION OF SOURCE IMAGES

| Category | Attribute Name | Pair Number |
|---|---|---|
| Characteristic | Edge | 59 |
| | Texture | 30 |
| | Symmetry | 6 |
| | Foreground | 38 |
| | Geometric structure | 24 |
| Environment | Indoor | 34 |
| | Outdoor | 66 |
| Content | Building | 32 |
| | People | 31 |
| | Nature | 37 |

categories include characteristic, environment, and content. And each category also contains various attributes, according to the acknowledged classification method [8]. The resolution of source images is 1920*1080 pixels. In our database, the resolution of retargeted images is 1344*1080 pixels. Our target is to analyze the perceptual quality of retargeted images. Therefore, without loss of generality, the shrinking ratio is set to 0.7 and we only make changes in the column direction.

In order to introduce the features about disparity, which are different from 2D images', we utilize the optical flow algorithm [13] to compute the disparity map of each stereoscopic image pair. For stereoscopic retargeting methods, we employ stereo cropping, stereo seam carving [14], stereo scaling and stereo multi-operator [15]. Note that stereo cropping and stereo scaling belong to the depth-adapting algorithms, while stereo seam carving and stereo multi-operator are based on the depth-preserving principle.

Stereo cropping is the operation of cutting both sides of left and right view images. When the cropped areas are the same in two views, the disparity remains unchanged. But when the cropping point of one view image moves, the disparity will be adjusted. Therefore, we select 54 images in stereo cropping dataset to make the disparity of half of these images close to or away from the visual comfort zone [16] respectively, which is calculated as [-79.55, 79.55] pixel converted from [-1°, 1°] of visual angle defined as visual comfort zone [17] according to our experiment environment.

Based on 2D seam carving algorithm, stereo seam carving considers extra disparity information [18]. The stereo seam carving finds the seam with the smallest energy to delete in the left view, and then removes the matched seam in the right view according to the disparity map. Thus this operation can preserve the same depth information as in original images.

As for stereo scaling, the method performs the same zoom operation on left and right views. There is no doubt that the distance between the corresponding pixels in two views is scaled, therefore the disparity also occurs in the same degree of scaling.

Stereo multi-operator chooses the best operator for per 36 columns from stereo cropping, stereo seam carving, and stereo scaling three operators. The decision of the best operator is made by comparing the defined energy function. It is worth noting here that the stereo cropping cuts equal parts in two view images, thus stereo multi-operator is a depth-preserving method.

The performance of above four image retargeting operators is presented in Fig. 1, where only left view images are shown.

We can find that stereo seam carving and stereo scaling introduce local geometric structure distortion, and stereo cropping generates global context information loss [19]. And stereo multi-operator may introduce both of the two artifacts.

*B. Subjective Experiment*

In our database, the subjective quality aspects contain visual comfort, image quality, depth quality and the overall quality. We follow the ITU-R BT.2021 [2] standard, and further customize them to assess these four aspects using single stimulus method on 5 discrete scales with 1 for bad quality and 5 for excellent quality. The viewing duration is 5s for each image. A total of 28 non-expert subjects who have passed the visual acuity, color blindness and stereo acuity tests, participated in the experiment. Their ages range from 17 to 28. Among the subjects, there are 21 males and 6 females. And the subjects rate on all these aspects in four separate tests.

For each pair of stereoscopic images, we utilize the 25 groups of subjective scores to calculate the corresponding Mean Opinion Score (MOS) after outlier removal. Also, we compute the correlation between the subjective rating of each viewer and the remaining 24 subjective scores. Further, the average, minimum and maximum of 25 correlation values are shown in Table II. From the table, we can see that the average correlation coefficients of the four quality assessment aspects represent a good consistency of subjective scores distribution and it can be seen that depth perception shows more diversity compared with other quality dimensions.

### III. PROPOSED VISUAL COMFORT METRIC

As shown in Fig. 2, we propose the VCA-SIR algorithm considering several impacts on visual comfort, e.g. visual comfort zone, window violation, binocular rivalry, accommodation-vergence (A/V) conflict adjustment intensity, and image quality, which are introduced by stereoscopic retargeting operations. The proposed metric contains four major features reflecting these impacts, which are disparity range, boundary disparity, disparity intensity distribution, and image quality features. Among them, image quality is generated from the perspective of 2D comfort evaluation. The other three features are designed from different visual symptoms in 3D perspective. Then these features are pooled with support vector regression (SVR) method, in which we utilize MOS of visual discomfort as labels.

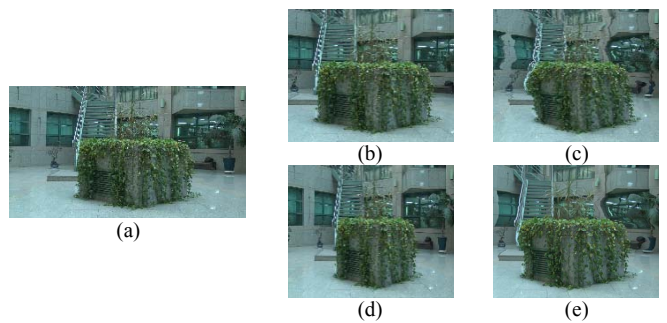

Fig.1 The left view images of a source and four retargeted stereoscopic images. (a) Source image, (b) Stereo cropping image, (c) Stereo seam carving image, (d) Stereo scaling image, (e) Stereo multi-operator image.

TABLE II. THE CORRELATION IN EACH QUALITY ASPECT

| Aspect | Visual Comfort | Image Quality | Depth Quality | Overall Quality |
|---|---|---|---|---|
| Average | **0.8223** | **0.8728** | **0.7414** | **0.8622** |
| Minimum | 0.7332 | 0.8106 | 0.6020 | 0.7990 |
| Maximum | 0.8754 | 0.9252 | 0.8481 | 0.9045 |

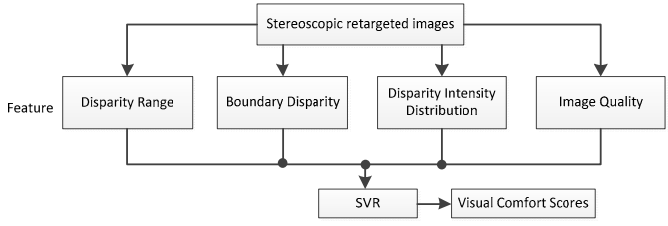

Fig.2 The diagram of VCA-SIR algorithm

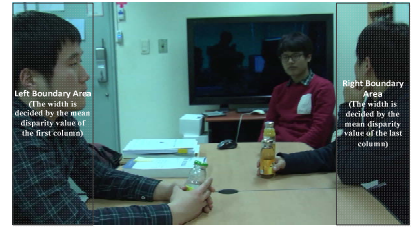

Fig.3 The definition of image boundary

*A. Disparity Range Feature*

When the disparity range of an image is too wide, it will be difficult for human eyes to fuse left and right views. Besides, the A/V conflict will occur [20]. This kind of conflict is a continuous switching process where both accommodate to focus on the display screen and concentrate on the object in order to perceive the virtual depth information. Also, if it cannot achieve a stable balance, the eyes will be very tired and uncomfortable. Obviously, it is necessary to ensure that the disparity range of the image is within the acceptable range, which is called comfort zone. In this case, viewers can have a relatively comfortable feeling when watching stereoscopic content.

We use the difference between the disparity range and the comfortable fusion zone as a measurement of comfort. The proposed disparity range feature (DR) is defined as:

$$R_r = \alpha \frac{d_{min}-x}{x} + \beta \frac{d_{max}-y}{y} \quad (1)$$

where $[x, y]$ is the disparity range of a stereoscopic image, $[d_{min}, d_{max}]$ is the visual comfort zone, which is $[-79.55, 79.55]$ pixel in our experiment. $\alpha$ and $\beta$ represent the penalty factor of the minimum and maximum disparity value beyond the comfort zone respectively, and satisfy $\alpha + \beta = 1$. According to [21], MT neurons' preferred disparity is crossed. In other words, human is more sensitive to crossed disparity than uncrossed. Thus, we set $\alpha = 0.4, \beta = 0.6$ in our experiment.

*B. Boundary Disparity Feature*

Stereoscopic retargeting operation may change the spatial position of foreground objects into the boundary area which is demonstrated in Fig. 3. In this case, the disparity in the boundary area will be larger than that in other areas since foreground objects usually have relatively larger disparity. The larger disparity then causes three phenomena which have an impact on visual comfort. First, the larger disparity in the boundary area is then more likely to result in more information which cannot be matched in left and right view images. Second, window violation can be induced due to the larger disparity and the visual field limitation [22]. Finally, the foreground objects located in the boundary area cannot be fully displayed at times, which tends to produce an incomplete sense of content.

At the same time, for other areas of an image, we consider whether the binocular rivalry occurs. When the left and right view stimulus are quite different, binocular rivalry happens. Then, human perception alters between left and right view visual signals. Some of stereoscopic retargeting operators, like stereo cropping, stereo scaling, are depth-adjusting, which can result in large disparity in the retargeted images. The large disparity then leads to binocular rivalry. Motivated by existing vision studies in this area [23], we use the variance of gray values to define the energy of each image. Then, we calculate the difference between the energy of left and right view images. It is thought that binocular rivalry will occur if the difference is large enough.

In a word, we calculate the average disparity of the boundary area to represent the probability of window violation and the completeness of image content, and calculate the difference between the energy of two view images to reflect the possibility of binocular rivalry. Let the size of the image is m*n, and the gray and disparity values on the $i^{th}$ row, the $j^{th}$ column of an image are $v_{ij}$ and $d_{ij}$, respectively. The left and right boundaries are defined as follows:

$$b_l = \frac{\sum_{i=1}^{n} d_{i1}}{n} \ and \ b_r = \frac{\sum_{i=1}^{n} d_{im}}{n} \quad (2)$$

Then, the average disparities of the left and right boundary areas are calculated as:

$$A_l = \frac{\sum_{j=1}^{b_l}\sum_{i=1}^{n} d_{ij}}{b_l \cdot n} \ and \ A_r = \frac{\sum_{j=m-b_r}^{m}\sum_{i=1}^{n} d_{ij}}{b_r \cdot n} \quad (3)$$

The energy of an image can be obtained as:

$$E = \frac{\sum_{j=1}^{m}\sum_{i=1}^{n}(v_{ij}-\bar{v})^2}{m*n} \quad (4)$$

where $\bar{v}$ is the mean of gray values in the image. The difference is defined as:

$$D = \frac{E_l}{E_r} \quad (5)$$

where $E_l$ and $E_r$ are the energy of left and right view image, respectively. Then, we form a three-dimensional feature vector $[A_l, A_r, D]$ as boundary disparity feature (BD).

*C. Disparity Intensity Distribution Feature*

As we can see from the above, the disparity range feature reflects the possibility of A/V conflict which is the main factor of visual discomfort [24]. The disparity intensity distribution feature (DID) proposed in this section will reflect the adjustment intensity of A/V conflict, that is, the amplitude of disparity change in one image. Imagine that if the amplitude of disparity change from one area to another area is too large, the steady state between accommodation and vergence has to reestablish, and the new steady state will have a significant difference from the previous one [25]. In other words, there is obvious hopping in depth. The more obvious the hopping is, the more discomfort will be felt. At this time, viewers have to focus on a specific point or object with big hopping for a short time to fit the new virtual depth by A/V adjustment. Besides, when viewers just watch the image with many disparity changes of large amplitude at first sight, they will feel dazzled. Therefore, it is necessary for VCA to introduce the DID feature.

The diagram of extracting DID feature is presented in Fig. 4. We first divide the entire image into 3*3 patches, and then classify the center pixel of each patch based on the grading ideas of Just Noticeable Depth Difference (JNDD), which is defined as [26]:

$$D_{JND} = \begin{cases} 21, & if\ 0 < |d_{ij}| < 64 : Bin1 \\ 19, & if\ 64 \leq |d_{ij}| < 128 : Bin2 \\ 18, & if\ 128 \leq |d_{ij}| < 192 : Bin3 \\ 20, & if\ 192 \leq |d_{ij}| < 255 : Bin4 \end{cases} \quad (6)$$

In the same way, according to JNDD, the other eight pixels in the patch are graded based on the center pixel. Thus, we obtain the disparity rank map. The reason of ranking is that the perceived disparity is limited. For example, for a reference point, whose surrounding disparity changes slightly, it is not easy to detect the reference point. While only when the surrounding disparity changes dramatically, it will be perceived [27]. Next, we calculate the rank gradients of horizontal, vertical and diagonal direction in each patch, and synthesize the gradients as a general gradient. Finally, the mean and variance of gradients of all patches are computed.

Meanwhile, we still divide the image into 3*3 patches. We then directly calculate the general gradient and the final mean and variance instead of ranking the pixel. We further obtain the weighted sum of the two pairs of mean and variance separately which serve as the DID feature vector.

### D. Image Quality Feature

The four retargeting methods mentioned above will introduce image distortion more or less. Thus, it is necessary to consider image quality in visual comfort assessment. Image quality assessment algorithm can be divided into full-reference, reduced-reference and no-reference image quality. Usually, the full-reference metric can achieve a better performance than the other two metrics. However the no-reference metric is more practical since the reference information is usually unavailable. Therefore, we only consider the full-reference and no-reference algorithms in this paper.

For full-reference image quality (FIQ), we utilize the bi-directional natural salient scene distortion model (BNSSD) [19], which includes image natural scene statistics measurement, salient global structure distortion measurement, and bi-directional salient information loss measurement. For no-reference image quality (NIQ), we use the oriented gradients image quality assessment (OG-IQA) algorithm [28] which proposed gradient magnitude, relative gradient orientation and relative gradient magnitude features. The "relative" word means that a pixel is regarded as a reference point, with its local area adjacent to the pixel in conjunction with consideration when calculate the features.

We use the algorithms to extract features from left and right view images separately and combine the corresponding features in each view as image quality feature.

### IV. PERFORMANCE

3D visual comfort analysis on heterogeneous devices is an important work for future 3D and immersive visual experiencing assessment. Although there are already some 3D comfort assessment solutions, as the features of stereoscopic retargeting operations, none of them can be directly applied for 3D image retargeting scenarios. Therefore, we validate the effectiveness of the proposed novel features by comparing the performance of the traditional full-reference and no-reference image quality metrics, proposed novels features and their combinations.

The SIRD database is randomly divided into 80% and 20% for training set and test set respectively, and the MOSs of visual comfort are used for the labels. The mean of correlations in 100 iterations of cross validation on the combination of features are calculated as shown in Table III.

From Table III, we can observe that the combination of four features outperforms other solutions, especially for the full-reference case. And the proposed three disparity features, i.e. DR, BD and DID, can effectively represent human perception for stereoscopic retargeted images. Also, we can find that the introduction of image quality feature obviously improves the performance of VCA-SIR. While existing image quality assessment models for stereoscopic image retargeting cannot fully reflect the visual comfort of images.

### V. CONCLUSION

In this paper, we build a subjective assessment database SIRD which will be published online; and aiming at the characteristics of stereoscopic retargeting methods, we propose new disparity range, boundary disparity and disparity intensity distribution features and introduce image quality feature to constitute the objective VCA-SIR. And the experimental results show that the performance of VCA-SIR is well accordant with human visual comfort perception.

In the future, we will do more experiments and compare the performance of existing VCA algorithms for stereoscopic images with VCA-SIR on our database. In addition, apart from VCA in this paper, we can also propose an overall quality assessment metric for stereoscopic image retargeting by integrating image quality and depth quality.

TABLE III. PERFORMANCE OF VCA MODEL ON OUR DATABASE

| Algorithm | PLCC | SRCC | KRCC | RMSE |
|---|---|---|---|---|
| FIQ | 0.8334 | 0.8298 | 0.6377 | 0.5863 |
| NIQ | 0.7786 | 0.7722 | 0.5777 | 0.6601 |
| DR+BD+DID | 0.8292 | 0.8107 | 0.6107 | 0.5841 |
| FIQ+DR+BD+DID | **0.9024** | **0.8958** | **0.7241** | 0.4544 |
| NIQ+DR+BD+DID | **0.8670** | **0.8494** | **0.6618** | 0.5237 |

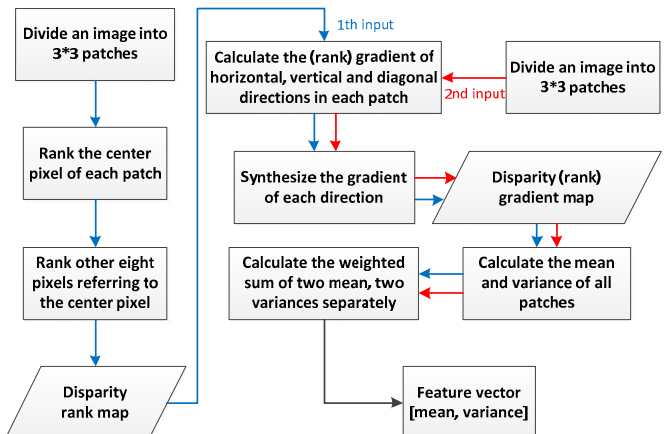

Fig.4 The extracting diagram of disparity intensity distribution feature.